\documentclass[roman,11pt]{article}

\usepackage{amsmath,amssymb,amsthm,amscd,latexsym}
\usepackage{times}
\usepackage{hyperref}
\usepackage{cite}
\usepackage{w-greek,wasysym,graphicx}

\def\rfr#1{eq. (\ref{#1})}

\def\eqi{\begin{equation}}
\def\eqf{\end{equation}}
\def\rp#1#2{{#1\over#2}} \def\lb#1{\label{#1}}
\def\bds#1{\boldsymbol{#1}}
\def\ton#1{\left(#1\right)}
\def\qua#1{\left[#1\right]}
\def\grf#1{\left\{#1\right\}}
\def\ang#1{\left\langle #1\right\rangle}
\def\nk{n_\mathrm{b}}

\begin{document}

\title{Might the 2PN perihelion precession of Mercury become measurable in the next future?}

\author{L. Iorio\\ Ministero dell'Istruzione, dell'Universit$\grave{\textrm{a}}$ e della Ricerca (M.I.U.R.)-Istruzione \\ Fellow of the Royal Astronomical Society (F.R.A.S.)\\ Viale Unit$\grave{\textrm{a}}$ di Italia 68, 70125, Bari (BA), Italy}

\maketitle

\begin{abstract}
The Hermean average perihelion rate $\dot\omega^\mathrm{2PN}$, calculated to the second post-Newtonian (2PN) order with the Gauss perturbing equations and the osculating Keplerian orbital elements, ranges from $-18$ to $-4$ microarcseconds per century $\ton{\mu\mathrm{as\,cty}^{-1}}$, depending on the true anomaly at epoch $f_0$. It is the sum of four contributions: one of them is the direct consequence of the 2PN acceleration entering the equations of motion, while the other three are indirect effects of the 1PN component of the Sun's gravitational field. An evaluation of the merely formal uncertainty of the experimental Mercury's perihelion rate $\dot\omega_\mathrm{exp}$ recently published by the present author, based on 51 years of radiotechnical data processed with the EPM2017 planetary ephemerides by the astronomers E.V. Pitjeva and N.P. Pitjev, is $\sigma_{\dot\omega_\mathrm{exp}}\simeq 8\,\mu\mathrm{as\,cty}^{-1}$, corresponding to a relative accuracy of $2\times 10^{-7}$ for the combination $\ton{2 + 2\gamma - \beta}/3$ of the PPN parameters $\beta$ and $\gamma$ scaling the well known 1PN perihelion precession.
In fact, the realistic uncertainty may be up to $\simeq 10-50$ times larger, despite reprocessing the now available raw data of the former MESSENGER mission with a recent improved solar corona model should ameliorate our knowledge of the Hermean orbit. The BepiColombo spacecraft, currently en route to Mercury, might reach a $\simeq 10^{-7}$ accuracy level in constraining $\beta$ and $\gamma$ in an extended mission, despite $\simeq 10^{-6}$ seems more likely according to most of the simulations currently available in the literature. Thus, it might be that in the not too distant future it will be necessary to include the 2PN acceleration in the Solar System's dynamics as well.
\end{abstract}



\centerline
{PACS: 04.80.-y; 04.80.Cc}

\section{Introduction}\lb{sec1}\lb{sec1}
The post-Newtonian (PN) approximation (see, e.g., \cite{1987thyg.book..128D,1997PThPS.128..123A,2003trso.conf..411B,2006LRR.....9....4B,2007LRR....10....2F,2018tegp.book.....W} and references therein) is a computational scheme for solving the Einstein's field equations of his General Theory of Relativity (GTR) relying upon the assumptions that  the characteristic speeds of the bodies under consideration are smaller than the speed of light $c$ and that the gravitational fields inside and around them are weak. Nonetheless, as pointed out in
\cite{2011PNAS..108.5938W}, such a scheme turned out to be remarkably effective in describing also certain strong-field and fast motion systems such as compact binaries made of at least one dense neutron star and inspiralling pairs of black holes emitting gravitational waves; the reasons for that are largely unknown \cite{2011PNAS..108.5938W}. Thus, putting the PN approximation to the test in as many different scenarios and at the highest order of approximation as possible is of paramount importance to gain an ever increasing confidence in it.

In its technical realm of validity, the PN approximation has been successfully tested so far only to the first post-Newtonian (1PN) order with, e.g., the orbital motions of planets, asteroids and spacecraft in the Solar System \cite{1968PhRvL..20.1517S,1972PhRvL..28.1594S,1971AJ.....76..588S,1990grg..conf..313S,2010PhRvL.105w1103L,2014PhRvD..89h2002L}, being its 2PN effects deemed too small to be currently measurable. The 1PN precession of the pericentre $\omega$ was measured also with binary pulsars \cite{2006Sci...314...97K,2021PhRvX..11d1050K} and stars orbiting the supermassive black hole at the centre of the Galaxy in Sgr A$^\ast$ \cite{2020A&A...636L...5G}.

The 2PN  precession of the pericentre of a two-body system, recently worked out\footnote{Among the several calculation existing in the literature with different computational schemes and parameterizations like, e.g., \cite{1993PhLA..174..196S,1995CQGra..12..983W,2019CQGra..36k5001T}, see \cite{1988NCimB.101..127D} for a derivation based on the Hamilton-Jacobi method and the Damour-Deruelle parameterization \cite{1985AIHS...43..107D}.} \cite{2021Univ....7..443I} in a perturbative way with the Gauss equations \cite{2011rcms.book.....K} and the standard osculating Keplerian orbital elements \cite{1994ApJ...427..951K}, was investigated in \cite{2020MNRAS.497.3118H} for the double pulsar PSR J0737-3039A/B \cite{2003Natur.426..531B,2004Sci...303.1153L} since it is viewed as a major source of systematic error in the expected future determination of the spin-orbit Lense-Thirring periastron precession \cite{1988NCimB.101..127D} since it should fall within the envisaged sensitivity level.

Here, we explore the perspectives of including, a day, the 2PN effects in the dynamics of the Solar System by looking at the perihelion of Mercury and the present and future level of accuracy in knowing its orbit. For the light propagation to the $\mathcal{O}\ton{c^{-4}}$ order and its possible astrometric measurements in the Solar System, see, e.g., \cite{2015IJMPD..2450056D} and references therein. Also various consequences of modified models of gravity were investigated to the 2PN order; see, e.g., \cite{2008PhRvD..77l4049X,2009AdSpR..43..171X,2012PhRvD..86d4007D,2015Deng,2016IJMPD..2550082D}.

The paper is organized as follows. In Section\,\ref{sec2}, we review the calculational strategy put forth in \cite{2021Univ....7..443I} to obtain the total 2PN precession of the pericenter of a two-body system. The results are applied in Section\,\ref{sec3} to Mercury and compared to the latest figures for the uncertainty in determining its perihelion rate. The role of the ongoing BepiColombo mission in improving the Hermean orbit determination is discussed as well. Section\,\ref{sec4} summarizes our findings and offers our conclusions.
\section{The 2PN precession of the pericenter}\lb{sec2}
The \textit{total} 2PN net\footnote{Here and in the following, the angular brackets $\ang{\ldots}$ denoting the orbital average are omitted.}  \textit{precession} of the argument of pericenter $\omega$ of a binary system made of two static, spherically symmetric bodies A and B, written in terms of the usual osculating Keplerian orbital elements, is \cite[eq.\, (18),\,pag.\,4]{2021Univ....7..443I}
\begin{align}
\dot\omega^\mathrm{2PN} \nonumber \lb{o2PN}& = \rp{3\,\mu^{5/2}}{8\,c^4\,a^{7/2}\,\ton{1-e^2}^3}\,\grf{
-68 + 8\,\nu + e^4\,\ton{-26 + 8\,\nu} + \right.\\ \nonumber \\
\nonumber &\left. + 2\,e^2\,\ton{-43 + 52\,\nu} + e\,\qua{8\,\ton{-29 + 13\,\nu} + e^2\,\ton{-8 + 61\,\nu}}\,\cos f_0  + \right.\\ \nonumber \\
&\left. +  3\,e^2\,\qua{4\,\ton{-5 + 4\,\nu}\,\cos 2 f_0  + e\,\nu\,\cos 3 f_0}}.
\end{align}
where $\mu\doteq G\,M$ is the gravitational parameter of the two-body system given by the product of the sum of its masses $M\doteq M_\mathrm{A} + M_\mathrm{B}$ by the Newtonian constant of gravitation $G$, the dimensionless parameter $\nu$ is given by $\nu\doteq M_\mathrm{A}\,M_\mathrm{B}/M^2$, $a,\,e$,  and $f_0$ are the osculating values of the semimajor axis, eccentricity and true anomaly, respectively, at the same arbitrary moment of time $t_0$ \cite{1994ApJ...427..951K}.

By expanding \rfr{o2PN} in powers of the eccentricity $e$, one gets
\begin{align}
\dot\omega^\mathrm{2PN} \nonumber \lb{oappr} &\simeq \rp{3\,\mu^{5/2}\,\ton{-17 +2\,\nu}}{2\,c^4\,a^{7/2}}+ \rp{3\,\mu^{5/2}\,\ton{-29 + 13\,\nu}\,\cos f_0}{c^4\,a^{7/2}}\,e + \\ \nonumber \\
& + \rp{3\,\mu^{5/2}\,\qua{-145 + 64\,\nu + 6\,\ton{-5 + 4 \nu}\,\cos 2 f_0}}{4\,c^4\,a^{7/2}}\,e^2 + \mathcal{O}\ton{e^3}.
\end{align}
From \rfr{oappr} it can be noted that, to the order of zero in $e$, the 2PN pericentre precession is independent of $f_0$.

It should be recalled that \rfr{o2PN} does not come only from the \textit{direct}\footnote{It is calculated perturbatively in the usual way by evaluating the right hand side of the Gauss equation for $\mathrm{d}\omega/\mathrm{d}t$ \cite{2011rcms.book.....K}, calculated with ${\bds A}^\mathrm{2PN}$, onto a \textit{fixed} Keplerian ellipse, and integrating it over a \textit{Keplerian} orbital period.} effect of the 2PN acceleration ${\bds A}^\mathrm{2PN}$ entering the equation of motion (see, e.g., \cite[eq.\,(38), pag.\,8]{2021Univ....7..443I}). Indeed, also two \textit{indirect}, or \textit{mixed}, effects  related to the 1PN acceleration ${\bds A}^\mathrm{1PN}$ (see, e.g., \cite[eq.\,(37), pag.\,7]{2021Univ....7..443I}) subtly concur to yield the net \textit{shift per orbit} $\Delta\omega$ to the 2PN level from which (a part of) the \textit{precession} follows by dividing it by the (Keplerian) orbital period  $P_\mathrm{b}$; see \cite{2015IJMPD..2450067I} for how to calculate such \textit{indirect} effects in a different scenario. They are due to the following facts. On the one hand, during an orbital revolution,  all the orbital elements entering the right hand side of the Gauss equation for $\mathrm{d}\omega/\mathrm{d}f$ \cite{1979AN....300..313M}, calculated with ${\bds A}^\mathrm{1PN}$, in principle, undergo instantaneous variations due to ${\bds A}^\mathrm{1PN}$ itself changing their values with respect to their fixed Keplerian ones referred to some reference epoch $t_0$. On the other hand, when the integration over $f$ is performed calculating  $\Delta\omega$, the fact that the true anomaly is reckoned from a generally varying line of apsides because of ${\bds A}^\mathrm{1PN}$ should be taken into account as well.
Such features yield two additional corrections to $\Delta\omega$ with respect to the usual 1PN one\footnote{It is obtained by keeping $a$ and $e$ \textit{fixed} during the integration of the right hand side of the Gauss equation for $\mathrm{d}\omega/\mathrm{d}f$, calculated with ${\bds A}^\mathrm{1PN}$,  over a \textit{Keplerian} orbital period.}
\eqi
\Delta\omega^\mathrm{1PN} = \rp{6\,\uppi\,\mu}{c^2\,a\,\ton{1-e^2}}\lb{1PN}
\eqf
which, in the case of ${\bds A}^\mathrm{1PN}$, are just of the order of $\mathcal{O}\ton{c^{-4}}$. Finally, as shown in \cite{2021Univ....7..443I}, when the \textit{total} 2PN net \textit{precession} has to be calculated, the 1PN \textit{fractional shift per orbit} $k^\mathrm{1PN}\doteq\Delta\omega^\mathrm{1PN}/2\uppi$ must be multiplied by the
1PN mean motion $\nk^\mathrm{1PN}$, and an expansion in powers of $c^{-1}$ to the order of $\mathcal{O}\ton{c^{-4}}$ must be taken.
Thus, the precession of \rfr{o2PN} comes from the sum of the latter contribution plus the \textit{direct} rate induced by ${\bds A}^\mathrm{2PN}$ and the two \textit{indirect} terms due to ${\bds A}^\mathrm{1PN}$.

In \cite{2021Univ....7..443I}, it is shown that \rfr{o2PN} agrees with other calculations existing in the literature performed with different computational strategies \cite{1988NCimB.101..127D,1994ARep...38..104K}. In particular, \rfr{o2PN} is in agreement with the expression for the total 2PN pericentre
precession, written in terms of the osculating Keplerian orbital elements, which can be inferred from \cite[eq.\,(5.18),\,pag.\,158]{1988NCimB.101..127D} based on the Damour-Deruelle parameterization \cite{1985AIHS...43..107D}.
\section{The 2PN perihelion precession of Mercury}
\begin{figure}[ht!]
\centering
\begin{tabular}{c}
\includegraphics[width = 12 cm]{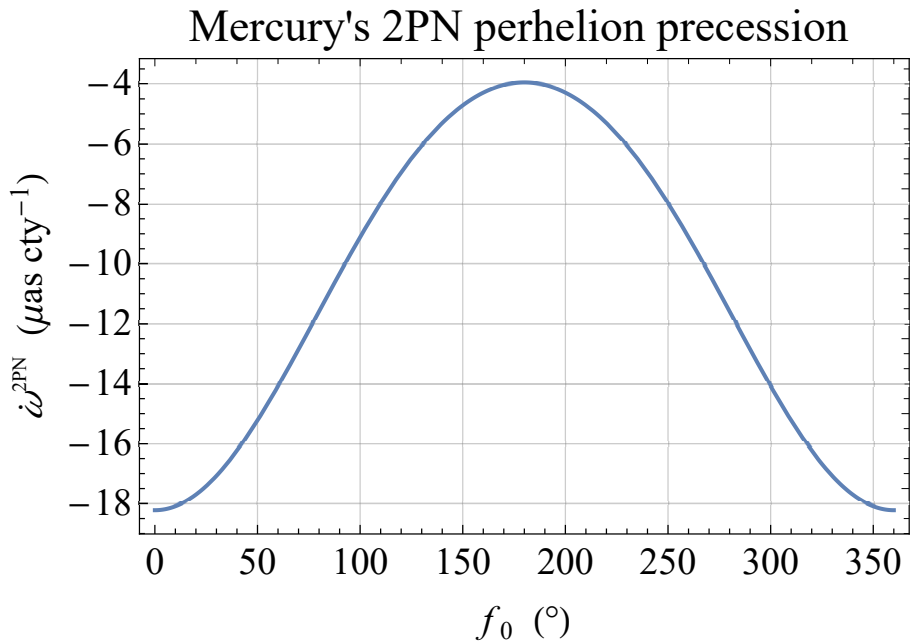}\\
\end{tabular}
\caption{
Total 2PN perihelion precession $\dot\omega^\mathrm{2PN}$ of Mercury, in  $\mu\mathrm{as\,cty}^{-1}$, as a function of the true anomaly at epoch $f_0$ according to \rfr{o2PN}. It turns out that $-18\,\mu\mathrm{as\,cty}^{-1}\lesssim \dot\omega^\mathrm{2PN}\lesssim -4\,\mu\mathrm{as\,cty}^{-1}$.
}\label{fig1}
\end{figure}

\lb{sec3}
In the case of Sun and Mercury\footnote{For a recent comparative study of Mercury's perihelion advance induced by some classical dynamical effects, see \cite{2022CeMDA.134...33P}.}, Figure\,\ref{fig1}, displaying the plot of \rfr{o2PN} as a function of $f_0$, shows that the Hermean 2PN precession ranges from $-18$ to $-4$ microarcseconds per century $\ton{\mu\mathrm{as\,cty}^{-1}}$.
About the current accuracy in determining observationally the perihelion precession $\dot\omega_\mathrm{exp}$ of Mercury, the present author in \cite{2019AJ....157..220I} tentatively inferred a \textit{formal} uncertainty as little as
\eqi
\sigma_{\dot\omega_\mathrm{exp}}\simeq 8\,\mu\mathrm{as\,cty}^{-1}\lb{erro}
\eqf
from the planetary ephemerides\footnote{EPM stands for Ephemeris of Planets and the Moon.} EPM2017 \cite{2018AstL...44..554P}. At first glance, \rfr{erro} might seem interesting since it is of the same order of magnitude of the 2PN contribution to the Hermean perihelion precession. The dynamical models of\footnote{See https://iaaras.ru/en/dept/ephemeris/epm/2017/ for details.} EPM2017 are accurate to the 1PN level, including, among other things, also the Lense-Thirring effect induced by the Sun's angular momentum $S_\odot$. As far as Mercury is concerned, they are based only on radio tracking data resulting in 1556 normal points over a temporal interval 51 years long (1964-2015); in particular, data collected by the MESSENGER  (Mercury Surface, Space Environment, Geochemistry and Ranging) spacecraft from 2011 to 2015  were analyzed. As pointed out in \cite{2018AstL...44..554P,2019AJ....157..220I}, realistic accuracies at the time of writing such papers may be $\simeq 10-50$ times larger than \rfr{erro}. On the other hand, a new model of solar plasma affecting the spacecraft ranging observations was recently published \cite{2022MNRAS.514.3191A}; it can now be used to reprocess the raw MESSENGER data\footnote{Until now,  the simpler model of  the NASA Jet Propulsion Laboratory (JPL) was pre-applied to the normal points published by it; see https://ssd.jpl.nasa.gov/dat/planets/messenger.txt} which were recently released\footnote{D. Pavlov, private communication to the present author, November 2022.}. This should improve the accuracy in our knowledge of the Mercury's orbit. As noted in \cite{2019AJ....157..220I}, \rfr{erro} corresponds to an uncertainty as little as $2\times 10^{-7}$ in the combination $\ton{2 + 2\,\gamma - \beta}/3$ of the PPN parameters $\gamma$ and $\beta$ in front of \rfr{1PN}. The ongoing mission to Mercury BepiColombo \cite{2021SSRv..217...90B} aims, among other things, to accurately determine $\beta$ and $\gamma$; according to \cite[Table\,5,\,pag.\,21]{2021SSRv..217...21I}, an extended mission may reach just the $\simeq 10^{-7}$ accuracy level in constraining such PPN parameters. The same conclusion is shown also in
\cite[Table\,2,\,pag.\,12]{2022RemS...14.4139V}; see also references therein. However, it should be remarked that most of the scenarios examined in \cite{2021SSRv..217...90B,2022RemS...14.4139V} envisage an accuracy in constraining $\beta$ and $\gamma$ of the order of $\simeq 10^{-6}$.
Be that as it may, perhaps, we may not be so far away from having to include, one day, also the 2PN terms in Solar System's dynamics.

In principle, a source of major systematic error which may overwhelm the 2PN perihelion precession is represented by the competing classical effect due to tidal distortion involving the Hermean Love number $k_2$ \cite{1911spge.book.....L,1959cbs..book.....K}. Such a parameter measures the mass concentration toward the centre of a fluid body assumed in hydrostatic equilibrium like, e.g., a main sequence star. Its possible values range from 0 for the mass point
approximation to $3/4=0.75$ for a fully homogeneous fluid body \cite{2014grav.book.....P}.
For a binary system, the periastron precession of tidal origin is \cite[eq.\,(3.100),\,pag.\,170]{2014grav.book.....P}
\eqi
\dot\omega_\mathrm{tid} = 15\,\nk\,\ton{1 + \rp{3}{2}\,e^2 + \rp{1}{8}\,e^4}\,\qua{k_2^\mathrm{A}\,\rp{M_\mathrm{B}}{M_\mathrm{A}}\,\ton{\rp{R_\mathrm{A}}{p}}^5 + \mathrm{A}\rightleftarrows\mathrm{B}},\lb{otid}
\eqf
where $p\doteq a\ton{1-e^2}$ is the semilatus rectum of the two-body relative orbit, and $R_\mathrm{A/B}$ is the equatorial radius of the body A or B.
A recent determination of the Love number of Mercury relying upon the analysis of the complete four years of MESSENGER tracking data from March 2011 to April 2015 yields\footnote{It has to be meant as the geophysicists' Love number, which is twice the astronomers' one, known also as apsidal constant, entering \rfr{otid} \cite[pag.\,115]{2014grav.book.....P}.} \cite{2020Icar..33513386K}
\eqi
k_2=0.53\pm 0.03\lb{k2}.
\eqf
The nominal value of the Hermean contribution\footnote{The one due to the Sun's Love number is much smaller, being, thus, of no concern.} to \rfr{otid}, calculated with \rfr{k2}, is of the order of a few $\mu\mathrm{as}\,\mathrm{cty}^{-1}$. Nonetheless, by assuming to model the tidal effects to the level of accuracy of \rfr{k2}, the resultant mismodeled perihelion precession would be well below the 2PN one.

From a practical point of view, experts in planetary data reductions should clarify which part of \rfr{o2PN} could be, actually, measured and how. Indeed, given that the 1PN equations of motion are currently included in the dynamical force models of the softwares routinely adopted to process the data, it may be argued that the indirect components of \rfr{o2PN} should have already been measured along with the 1PN precession; if so, only the direct part due to ${\bds A}^\mathrm{2PN}$ could be detected by explicitly modeling it and estimating, say, a dedicated scaling solve-for parameter. Otherwise, ${\bds A}^\mathrm{2PN}$ should be modeled in terms of $\beta$ and $\gamma$ whose newly estimated values, if known at the $10^{-7}$ level, would account for the 2PN component of the equations of motion as well.

\section{Summary and conclusions}\lb{sec4}
The 2PN pericentre precession $\dot\omega^\mathrm{2PN}$ of a two-body system made of two static, spherically symmetric masses, obtained with the standard Gauss perturbative equation in terms of the usual osculating Keplerian orbital elements, is reviewed. Both the exact expression (\rfr{o2PN}) and an approximated expansion in powers of the eccentricity $e$ up to the order of $\mathcal{O}\ton{e^2}$ (\rfr{oappr}) are given. It is recalled that $\dot\omega^\mathrm{2PN}$ consists of four contributions: a direct term straightforwardly arising from the 2PN acceleration ${\bds A}^\mathrm{2PN}$ in the equations of motion, and three indirect parts due to the 1PN acceleration ${\bds A}^\mathrm{1PN}$ itself. In particular, one of the latter ones arises from the expansion to the order of $\mathcal{O}\ton{c^{-4}}$ of the product of the well known fractional 1PN shift per orbit (the ratio of \rfr{1PN} to $2\,\uppi$) by the 1PN mean motion. The other two indirect contributions come from taking into account  also the instantaneous variations of the order of $\mathcal{O}\ton{c^{-2}}$  of the orbital elements and the fact that the anomalistic period over which the 1PN shift is integrated is the time interval between two successive crossings of an actually moving pericenter due to ${\bds A}^\mathrm{1PN}$ itself.  The resulting total 2PN precession of \rfr{o2PN} depends on the true anomaly at epoch $f_0$. It is remarked that \rfr{o2PN} agrees with other expressions for $\dot\omega^\mathrm{2PN}$ in the literature obtained with different parameterizations and calculational schemes.

In the case of the Sun and Mercury, the Hermean 2PN perihelion precession, calculated with \rfr{o2PN}, ranges from $-18$ to $-4\,\mu\mathrm{as\,cty}^{-1}$ depending on $f_0$. A recent guess for the formal experimental uncertainty in determining the Mercury's perihelion precession with the EPM2017 planetary ephemerides is $\sigma_{\dot\omega_\mathrm{exp}}\simeq 8\,\mu\mathrm{as\,cty}^{-1}$, although the realistic uncertainty may be up to $\simeq 10-50$ times larger. Nonetheless, the raw data collected during the past MESSENGER mission are now available, and a new, accurate model of the solar corona, usually a major bias impacting the accuracy of ranging measurements, was recently published. Thus, reprocessing the  MESSENGER observations  with such a new model should improve our knowledge of Mercury in the near future.  A $\simeq\mu\mathrm{as\,cty}^{-1}$ level corresponds to an uncertainty of the order of $\simeq 10^{-7}$ in the PPN combination $\ton{2+2\gamma-\beta}/3$
scaling the 1PN precession. In a few scenarios encompassing an extended mission profile of the BepiColombo spacecraft, currently en route to Mercury, such a level of accuracy in constraining the PPN parameters $\gamma$ and $\beta$ may be reached, although $\simeq 10^{-6}$  seems more plausible according to the majority of the simulations performed so far in the literature. Anyway, perhaps, the time when it will be necessary to model the dynamics of the Solar System at the 2PN order might not be that far away, after all.
\section*{Acknowledgements}
I am indebted to M. Efroimsky for his careful and attentive comments and remarks, and to R.S. Park for clarifications about the Love number of Mercury. 
I gratefully thanks also G. Tommei and D. Pavlov for useful information.
\bibliography{2PN.bib}{}

\end{document}